\documentclass[12pt]{article}

\usepackage{amssymb}
\usepackage{amsmath}
\usepackage{amscd}
\usepackage{latexsym}

\usepackage{graphicx}
\usepackage{array}

\usepackage{cite}

\newcommand{\be}{\begin{equation}}
\newcommand{\ee}{\end{equation}}
\newcommand{\Dlt}{\Delta}
\newcommand{\dlt}{\delta}

\newcommand{\br}{{\bf r}}

\newcommand{\bn}{{\bf n}}
\newcommand{\bfe}{{\bf e}}
\newcommand{\ba}{{\bf a}}

\newcommand{\bt}{\beta}
\newcommand{\vp}{\varphi}

\newcommand{\al}{\alpha}
\newcommand{\ra}{\rightarrow}

\newcommand{\gm}{\gamma}
\newcommand{\om}{\omega}
\newcommand{\Om}{\Omega}

\newcommand{\dgr}{\dagger}

\newcommand{\bB}{{\bf B}}
\newcommand{\bF}{{\bf F}}
\newcommand{\bS}{{\bf S}}

\newcommand{\rgl}{\rangle}
\newcommand{\lgl}{\langle}

\begin{document}

\begin{center}

{\Large{\bf Spin dynamics in lattices of spinor atoms with quadratic 
Zeeman effect} \\ [5mm]

V.I.~Yukalov$^{1,2,*}$ and E.P.~Yukalova$^{3}$ } \\ [3mm] 

{\it 
$^1$Bogolubov Laboratory of Theoretical Physics, \\
Joint Institute for Nuclear Research, Dubna 141980, Russia \\ [2mm]
                                           
$^2$Instituto de Fisica de S\~ao Carlos, Universidade de S\~ao Paulo, \\
CP 369,  S\~ao Carlos 13560-970, S\~ao Paulo, Brazil  \\ [2mm]

$^3$Laboratory of Information Technologies, \\
Joint Institute for Nuclear Research, Dubna 141980, Russia } \\ [1cm]

$^*$ Corresponding author
E-mail address:  yukalov@theor.jinr.ru (V.I. Yukalov)     

\end{center}

\vskip 1cm

\begin{abstract}
A lattice system of spinor atoms or molecules experiencing quadratic Zeeman effect 
is considered. This can be an optical lattice with sufficiently deep wells at lattice 
sites, so that the system is in an isolating state, where atoms are well localized. But 
their effective spins can move in the presence of external magnetic fields. The dynamics 
of spins, starting from an initial nonequilibrium state, is investigated. The system 
is immersed into a magnetic coil of an electric circuit, creating a magnetic feedback 
field. Two types of quadratic Zeeman effect are treated, a nonresonant, so-called 
static-current quadratic Zeeman effect and a quasi-resonant alternating-current quadratic 
Zeeman effect. Spin dynamics in these conditions is highly nonlinear. Different regimes 
of spin dynamics, starting from a strongly nonequilibrium state, are studied. Conditions 
for realizing fast spin reversal are found, which can be used in quantum information 
processing and spintronics.
\end{abstract}

\newpage

\section{Introduction}

Atomic (or molecular) systems interacting through dipolar and spinor forces have been 
a topic of intensive research in recent years, as can be inferred from the books and 
review articles 
\cite{Griesmaier_1,Baranov_2,Pethick_3,Ueda_4,Baranov_5,Gadway_6,Stamper_7,Yukalov_8,Yukalov_9}.
A great advantage of these systems is the richness of their properties and the possibility 
of tuning the latter in a rather wide range. 

In the present paper, we study spinor atomic systems forming lattices. These can be 
either self-organized lattices or optical lattices created by laser beams. Our main 
concern is not the motion of atoms, but the dynamics of effective spin variables. 
Therefore we consider deep lattices, where atoms, being well localized, form insulating 
states. Thus the atomic motion is frozen, while effective spins can move, especially if 
external magnetic fields are applied.

In the presence of external magnetic fields, spinor atoms can experience quadratic Zeeman 
effect. There are, actually, two types of the effect. One is the nonresonant static-current 
quadratic Zeeman effect arising in atoms possessing hyperfine structure, hence a nonzero 
nuclear spin \cite{Jenkins_10,Schiff_11,Killingbeck_12,Coffey_13}.     

The other type of quadratic Zeeman effect is the so-called quasi-resonant 
alternating-current quadratic Zeeman effect, due to the alternating-current Stark shift. 
The latter can be produced by applying either a linearly polarized microwave driving 
field inducing hyperfine transitions in an atom \cite{Gerbier_14,Leslie_15,Bookjans_16} 
or applying off-resonance linearly polarized light inducing transitions between internal 
spin states \cite{Cohen_17,Santos_18,Jensen_19,Paz_20}. The linearly polarized 
quasiresonance alternating fields exert the quadratic shift along the polarization axis.  

We shall take into account both these phenomena, the static-current quadratic Zeeman 
effect and the alternating-current quadratic Zeeman effect. Our aim is to study how the 
presence of these effects influences spin dynamics in spinor lattices and how this 
influence could be employed for governing spin motion.  
  
The main difference of the present work from the previous publications on spinor atomic 
systems is in the following five points:
\begin{enumerate}
\item
We concentrate our attention on the study of spin dynamics, while the spatial motion of 
atoms is frozen in a deep insulating lattice. This regime is important for spintronics.   

\item
We consider strongly nonequilibrium dynamics, but not slight deviations from equilibrium,
which requires to deal with special methods of solving spin evolution equations.

\item
Both types of the quadratic Zeeman effect are taken into account, the nonresonant 
static-current, as well as the quasi-resonant alternating-current effects, which 
provides efficient tools for influencing spin motion.    

\item
In addition to a stationary magnetic field, the sample is subject to the action of 
a feedback magnetic field formed by a magnetic coil of an electric circuit. Such a 
configuration allows for a powerful possibility of regulating spin dynamics.    

\item
Conditions for realizing fast spin reversal are investigated. This mechanism can be 
employed in spintronics and information processing. 
\end{enumerate}

\section{Hamiltonian of spinor atoms}

A spinor atom (or molecule) is characterized by a total angular momentum ${\bf F}$ that 
is a matrix vector
\be 
\label{1}
 \bF = [\bF_{mn}] = \sum_\al F^\al\bfe_\al = \sum_\al [F^\al_{mn} ]\bfe_\al 
\qquad
(\al = x,y,z) \;  ,
\ee
with the matrix elements labeled by the index $m = -F, -F + 1, \ldots, F$. The related 
field operators are the columns
\be
\label{2}
\hat\psi(\br) = [\hat\psi_m(\br) ] \qquad ( m = -F,-F+1,\ldots,F) \;  .
\ee
The time dependence of the field operators is assumed, but not shown for brevity. The 
total Hamiltonian of a system of spinor atoms can be written as a sum of three terms
\be
\label{3}
 \hat H = \hat H_0 + \hat H_Z + \hat H_{int} \; .
\ee
Here the first term is the single-atom part not containing the angular momentum,
\be
\label{4}
 \hat H_0 = \int \hat\psi^\dgr(\br) \left [ - \hbar^2 \; \frac{\nabla^2}{2m} +
U(\br) \right ]  \hat\psi(\br) \; d\br \; ,
\ee
where $U({\bf r})$ is an external field, for instance, due to an optical lattice.    

The second term is the Zeeman energy Hamiltonian
\be
\label{5}
\hat H_Z = \hat H_{LZ} + \hat H_{QZ}
\ee
including the parts caused by the linear and quadratic Zeeman effects. The linear 
Zeeman term is
\be
\label{6}
 \hat H_{LZ} = - \mu_F \int \hat\psi^\dgr(\br)\; \bB\cdot\bF\; \hat\psi(\br) \; d\br \; , 
\ee
where $\mu_F = - g_F \mu_B$, with $g_F$ being the Land\'{e} factor and $\mu_B$, Bohr 
magneton. The external magnetic field ${\bf B} = {\bf B}({\bf r}, t)$, in general, is 
a function of spatial and time variables. 

As is mentioned in the Introduction, quadratic Zeeman effect can be of two types, 
nonresonant static-current Zeeman effect and quasi-resonant alternating-current Zeeman 
effect. The corresponding Hamiltonian is
\be
\label{7}
\hat H_{QZ} = Q_Z \int \hat\psi^\dgr(\br)\; (\bB\cdot\bF)^2\; \hat\psi(\br)\;d\br \; + 
\; q_Z \int \hat\psi^\dgr(\br)\; (F^z)^2 \; \hat\psi(\br) \; d\br  \;  ,
\ee
with the static-current Zeeman parameter
\be
\label{8}
 Q_Z = \mp \; \frac{\mu_F^2}{\Dlt W(1+2I)^2} \;  ,
\ee
in which $\Delta W$ is a hyperfine energy splitting and $I$ is nuclear spin, and with 
the alternating-current Zeeman parameter
\be
\label{9}
q_Z = - \; \frac{\hbar \Om^2_R}{4\Dlt} \;   ,
\ee
where $\Omega_R$ is the Rabi frequency of the driving alternating field and $\Delta$ 
is the detuning from an internal (spin or hyperfine) transition. The polarization of 
the driving field is assumed to be along the axis $z$. The sign minus or plus in the 
static-current Zeeman parameter $Q_Z$ is defined by the relative alignment of the 
nuclear and the total electron spin projections of the atom: minus for parallel 
projections, while plus for antiparallel projections. The sign of the alternating-current 
Zeeman parameter $q_Z$ can be varied by using either positive or negative detuning. The 
parameter $q_Z$ can be tailored at high resolution and rapidly adjusted to the desired 
values. 

The interaction Hamiltonian
\be
\label{10}
\hat H_{int} = \hat H_F + \hat H_D
\ee
is the sum of a term $\hat{H}_F$ describing local interactions of atoms, each having an
angular momentum ${\bf F}$, and of a term $\hat{H}_D$ corresponding to nonlocal dipolar 
interactions. For rotationally symmetric pair collisions in the $s$-wave approximation,
the angular momentum $f$ of the pair of colliding atoms has to be even, which is valid
for bosons as well as for fermions. Then the binary collisions of atoms are characterized
by the interaction potential
\be
\label{11}
 \Phi_F(\br) = \dlt(\br) \sum_f 4\pi\hbar^2 \; \frac{a_f}{m} \hat P_f \;  ,
\ee
where $a_f$ is the scattering length of a pair of atoms with the angular momentum of 
the pair $f$ and $\hat{P}_f$ is a projection operator onto a state with an even angular 
momentum $f$. The Hamiltonian of such local atomic interactions reads as
\be
\label{12}
  \hat H_F = \frac{1}{2} \sum_{klmn}  \int \hat\psi^\dgr_k(\br) \hat\psi^\dgr_l(\br)
\Phi_{klmn} \hat\psi_m(\br) \hat\psi_n(\br) \; d\br \; ,
\ee
with $\Phi_{klmn}$ being a matrix element of potential (\ref{11}).     

Atoms, possessing angular momenta, also interact through dipolar forces. The related 
dipolar interaction Hamiltonian is
\be
\label{13}
 \hat H_D = \frac{\mu_F}{2} \sum_{klmn}  \int  \hat\psi^\dgr_k(\br) \hat\psi^\dgr_l(\br')
 \hat\psi_m(\br') \hat\psi_n(\br) D_{klmn}(\br-\br') \; d\br \; d\br' \;  ,
\ee
with the dipolar interaction potential 
\be
\label{14}
 D_{klmn}(\br) = 
\frac{(\bF_{kn}\cdot\bF_{lm} ) -3 (\bF_{kn}\cdot\bn)((\bF_{lm}\cdot\bn)}{r^3} \; ,
\ee
where $r \equiv |{\bf r}|$ and ${\bf n} \equiv {\bf r}/ r$. Strictly speaking, the 
dipolar interactions should be regularized by taking into account the sizes of atoms 
and, in general, screening effects \cite{Yukalov_8,Yukalov_9,Jonscher_21}. The 
regularized dipolar potential can be written in the form
\be
\label{15}
 \overline D_{klmn}(\br) = \Theta(r-b_F)  D_{klmn}(\br) \exp(-\varkappa_F r) \;  ,
\ee
in which $\Theta(r)$ is a unit-step function, $b_F$ is a short-range cutoff, and 
$\varkappa_F$ is a screening parameter. The regularized potential automatically 
excludes unphysical self-action, so that Hamiltonian (\ref{13}) can be represented 
as
\be
\label{16}
 \hat H_D = \frac{\mu_F}{2} \sum_{klmn}  \int  \hat\psi^\dgr_k(\br) \hat\psi^\dgr_n(\br)
\overline D_{klmn}(\br-\br')
 \hat\psi_l(\br') \hat\psi_m(\br') \; d\br \; d\br' \;  .
\ee
In what follows, we shall denote, for short, the dipolar potential as $D_{klmn}({\bf r})$, 
while keeping in mind its regularized form, when it is necessary to avoid unphysical 
consequences.

\section{Deep insulating lattice}

Suppose the atoms form a lattice, with the lattice vectors ${\bf a}_j$, where the index 
$j = 1,2, \ldots, N$ enumerates lattice sites. Resorting to the single-band approximation,  
the field operators can be expanded over Wannier functions,
\be
\label{17} 
 \hat\psi_m(\br) = \sum_j \hat c_{jm} w(\br-\ba_j) \;  .
\ee 
The Wannier functions are assumed to be independent of the hyperfine index and can be chosen
to be well localized \cite{Marzari_22}, so that the lattice is insulating and intersite 
tunneling can be neglected. 

It is possible to introduce effective spin operators
\be
\label{18}
\bS_j \equiv \sum_{mn} \hat c^\dgr_{jm} \bF_{mn} \hat c_{jn}
\ee
that are localized in the lattice sites. These operators satisfy the standard spin algebra
for any type of statistics of atoms, whether bosons or fermions. 

For a deep lattice, without tunneling, the total Hamiltonian (\ref{3}) can be written as
the sum 
\be
\label{19}
\hat H = \hat H_L + \hat H_S
\ee
of the local Hamiltonian 
\be
\label{20}
\hat H_L = \hat H_0 + \hat H_F
\ee
and of the effective spin Hamiltonian
\be
\label{21}  
\hat H_S = \hat H_Z + \hat H_D \;   .
\ee
The local Hamiltonian (\ref{20}) can be shown \cite{Yukalov_9} to commute with the spin 
Hamiltonian $H_S$, which we illustrate below. Therefore spin dynamics is governed only 
by the effective spin Hamiltonian. 

The local Hamiltonian depends on the hyperfine angular momentum of atoms. Thus for $F=1$,
the components of the angular momentum are
$$
F_{mn}^x = \frac{1}{\sqrt{2}} \; (\dlt_{m\;n-1} + \dlt_{m\;n+1} ) \; , \qquad
F_{mn}^y = \frac{i}{\sqrt{2}} \; (\dlt_{m\;n-1} - \dlt_{m\;n+1} ) \; ,
$$
$$
F_{mn}^z =  m\dlt_{mn} \qquad (m,n = -1,0,1) \; .
$$
As a result, we have 
\be
\label{22}
 \hat H_L =\sum_j \left [ h_j \hat n_j + 
\frac{\overline c_0}{2} \; \hat n_j (\hat n_j + 1)  + 
\frac{\overline c_2}{2} \; \left ( \bS_j^2 - 2\hat n_j \right ) \right ] \; ,
\ee
with the local-site energy
\be
\label{23}
h_j \equiv \int w^*(\br-\ba_j) \left [ - \hbar^2 \; \frac{\nabla^2}{2m} + 
U(\br) \right ] w(\br-\ba_j) \; d\br \; ,
\ee
the operator density of atoms
\be
\label{24}
  \hat n_j \equiv \sum_m \hat c^\dgr_{jm} \hat c_{jm} \; ,
\ee
and the interaction parameters
\be
\label{25}
 \overline c_0 \equiv c_0 \int | w(\br-\ba_j)|^4 \; d\br \; , \qquad
 \overline c_2 \equiv c_2 \int | w(\br-\ba_j)|^4 \; d\br \; ,
\ee
in which
\be
\label{26}
 c_0 \equiv \frac{4\pi}{3m} \; \hbar^2 ( 2a_2 + a_0) \; , \qquad 
c_2 \equiv \frac{4\pi}{3m} \; \hbar^2 ( a_2 - a_0) \;  .
\ee
Here $a_0$ and $a_2$ are the scattering lengths for the collisions of atoms with the
moments of atom pairs $f = 0$ and $f = 2$, respectively. 

If the external magnetic field varies in space slower than the variation of the well 
localized Wannier functions, then the Zeeman Hamiltonian (\ref{5}) takes the form
\be
\label{27}
 \hat H_Z = \sum_j \left [ - \mu_F \bB_j\cdot \bS_j + Q_Z (\bB_j\cdot\bS_j)^2 +
q_Z ( S_j^z)^2 \right ] \;  ,
\ee
where ${\bf B}_j \equiv {\bf B}({\bf a}_j)$. The dipolar Hamiltonian (\ref{13}), for 
well localized atoms, becomes
\be
\label{28}
 \hat H_D = \frac{1}{2} \sum_{i\neq j} \; 
\sum_{\al\bt} D_{ij}^{\al\bt} S_i^\al S_j^\bt \;   ,
\ee
with the dipolar tensor
\be
\label{29}
 D_{ij}^{\al\bt} = \frac{\mu_F^2}{r_{ij}} \; (\dlt_{\al\bt} - 3 n_{ij}^\al n_{ij}^\bt)
\exp(-\varkappa_F r_{ij} ) \;  ,
\ee 
in which
$$
r_{ij} \equiv |\br_{ij} | \; , \qquad \bn_{ij} \equiv \frac{\br_{ij}}{r_{ij} } \; ,
\qquad
 \br_{ij} \equiv \br_i - \br_j \; .
$$
For generality, we keep here the screening factor, which, although, is not principal 
for the following. If the screening is absent, then one should set $\varkappa_F = 0$. 
The short-range cutoff can be omitted here, since the summation in Eq. (\ref{28}) does 
not include coinciding lattice indices. A detailed derivation of Hamiltonian (\ref{19}) 
for atoms with $F = 1$ in a deep lattice can be found in the review \cite{Yukalov_9}. 
 
The external magnetic field consists of a constant field $B_0$ along the axis $z$ and 
of a feedback field $H$ along the axis $x$, so that
\be
\label{30}
 \bB_j = B_0\bfe_z + H \bfe_x \;  .
\ee 
For the following, it is convenient to use the ladder spin operators connected with 
the spin components 
$$
 S_j^x = \frac{1}{2} \; ( S_j^+ + S_j^- ) \; , \qquad
  S_j^y = -\; \frac{i}{2} \; ( S_j^+ - S_j^- ) \; .
$$
Then the Zeeman Hamiltonian reads as
$$
\hat H_Z = \sum_j \left\{ - \mu_F B_0 S_j^z - \; \frac{\mu_F H}{2} \; ( S_j^+ + S_j^- )
+ ( Q_Z B_0^2 + q_Z) ( S_j^z)^2 \; + \right.
$$
$$
+ \; 
\frac{Q_Z H^2}{4} \; \left [ ( S_j^+)^2  + ( S_j^-)^2 + S_j^+ S_j^- + S_j^- S_j^+ 
\right ] \; +
$$
\be
\label{31}
 \left. 
+ \;
\frac{Q_Z B_0 H}{2} \; \left [ ( S_j^+ + S_j^- )S_j^z + S_j^z ( S_j^+ + S_j^- ) \right ]
\right \} \; .
\ee
And the dipolar term takes the form
\be
\label{32}
 \hat H_D = \frac{1}{2} \sum_{i\neq j} \left [ a_{ij} \left ( S_i^z S_j^z - \; 
\frac{1}{2} \; S_i^+  S_j^- \right ) + b_{ij} S_i^+  S_j^+ + b_{ij}^* S_i^-  S_j^- 
+2c_{ij} S_i^+  S_j^z + 2c_{ij}^* S_i^-  S_j^z \right ] \; ,
\ee
in which
$$
 a_{ij} \equiv D_{ij}^{zz} \; , \qquad b_{ij} \equiv \frac{1}{4} \; ( D_{ij}^{xx} -
D_{ij}^{yy} - 2i D_{ij}^{xy} ) \; , \qquad
c_{ij} \equiv \frac{1}{2} \; ( D_{ij}^{xz} - i D_{ij}^{yz} ) \;  .
$$

Note that the short-range regularization, excluding self-action, implies that
$$
D_{jj}^{\al\bt} = 0 \; , \qquad a_{jj} = b_{jj} = c_{jj} = 0 \;  .
$$

\section{Spin equations of motion}

The constant external magnetic field defines the Zeeman frequency
\be
\label{33}
 \omega_0 \equiv - \mu_F B_0 > 0 \;  .
\ee
To simplify the following formulas, we introduce the effective quadratic Zeeman parameter
\be
\label{34}
 Q \equiv Q_Z B_0^2 + q_z \;  ,
\ee
define the local spin fluctuating fields
\be
\label{35}
\xi_i \equiv \frac{1}{\hbar} 
\sum_j ( a_{ij} S_j^z + c_{ij} S_j^+ + c_{ij}^* S_j^- ) \; , 
\qquad  
\vp_i \equiv \frac{1}{\hbar} 
\sum_j \left ( \frac{a_{ij}}{2} S_j^- -2 b_{ij} S_j^+ - 2c_{ij} S_j^z \right ) \;,
\ee
and use the notation
\be
\label{36}
f_j \equiv - i \left ( \frac{\mu_F H}{\hbar} + \vp_j \right ) \;  .
\ee

Employing the Heisenberg equations of motion, we find the equation for the ladder spin 
operator  
$$
\frac{dS_j^-}{dt} = - i ( \om_0 + \xi_j) S_j^- + f_j S_j^z - 
\; \frac{i}{\hbar}\; Q ( S_j^- S_j^z + S_j^z S_j^- ) \; +
$$  
$$
+\;
\frac{i}{2\hbar}\; Q_Z H^2 
\left [ ( S_j^+ + S_j^- ) S_j^z + S_j^z ( S_j^+ + S_j^- ) \right ] \; -
$$
\be
\label{37}
 - \;
\frac{i}{2\hbar}\; Q_Z B_0 H \left [  S_j^+ S_j^- + S_j^- S_j^+ - 4(S_j^z)^2 +
2 ( S_j^-)^2 \right ]
\ee
and the equation for the longitudinal spin component
$$
\frac{dS_j^z}{dt} = - \; \frac{1}{2} ( f_j^+ S_j^- + S_j^+ f_j ) \; +
$$
\be
\label{38}
 + \; \frac{i}{2\hbar}\; Q_Z H^2 
\left [ ( S_j^-)^2 - (S_j^+)^2 \right ] +
\frac{i}{2\hbar}\; Q_Z B_0 H 
\left [ ( S_j^- - S_j^+ ) S_j^z + S_j^z ( S_j^- - S_j^+ ) \right ] \; .
\ee

From these equations, we can derive the equations for the statistical averages of spin 
components. We are looking for the equations for the following quantities: the average
transverse spin variable
\be
\label{39}
u = \frac{1}{SN} \sum_j \; \lgl S_j^- \rgl \;   ,
\ee
the coherence intensity
\be
\label{40}
w = \frac{1}{SN(N-1)} \sum_{i\neq j} \; \lgl S_i^+ S_j^- \rgl \;   ,
\ee
and the longitudinal spin polarization
\be
\label{41}
 s = \frac{1}{SN} \sum_j \; \lgl S_j^z \rgl \;  .
\ee

To get a complete set of equations, we need to decouple the spin correlation functions.
For different lattice sites, we use the mean-field approximation
\be
\label{42}
 \lgl S_i^\al S_j^\bt \rgl = \lgl S_i^\al  \rgl \lgl S_j^\bt \rgl 
\qquad  (i \neq j) \; .
\ee
But for binary spin forms at the same site this approximation cannot be used, since, 
for instance, when $S = 1/2$, then there is the exact equality
$$
 S_j^\al S_j^\bt + S_j^\bt S_j^\al = 0 \qquad \left ( S = \frac{1}{2} \right ) \; ,
$$
while the standard mean-field approximation would result in a nonzero quantity. The
correct approximation for such single-site binary combinations, for arbitrary spins,   
reads \cite{Yukalov_23,Yukalov_24,Yukalov_25} as
\be
\label{43}
\lgl  S_j^\al S_j^\bt + S_j^\bt S_j^\al \rgl =  \left ( 2 -\; \frac{1}{S} \right )
\lgl S_j^\al  \rgl \lgl S_j^\bt  \rgl \; .
\ee
This decoupling is exact for $S = 1/2$ and asymptotically exact for large spins. 
 
The other difficulty is that, averaging the terms of the type $\sum_j \xi_j S^\al_j$,
for an ideal lattice, in the mean-field approximation one gets zero, because of the 
property of the dipolar tensor
$$
 \sum_j D_{ij}^{\al\bt} = 0 \; , \qquad \sum_j a_{ij} = \sum_j b_{ij} =
\sum_j c_{ij}  = 0 \;  .
$$
A more refined method, retaining the input of local dipolar fluctuations, is based on 
stochastic quantization \cite{Yukalov_26,Yukalov_27}. For this purpose, we define the 
averages
\be
\label{44}
\left \lgl \frac{1}{N} \sum_j \xi_j S_j^\al \right \rgl = 
\xi_S \; \frac{1}{N} \sum_j \; \lgl S_j^\al \rgl \; ,
\qquad 
\left \lgl \frac{1}{N} \sum_j \vp_j S_j^\al \right \rgl = 
\vp_S \; \frac{1}{N} \sum_j \; \lgl S_j^\al \rgl \; .
\ee
The variables $\xi_S$ and $\varphi_S$, describing local dipolar fluctuations, are 
treated as stochastic variables. It is straightforward to show that these variables 
are responsible for the existence of dipolar spin waves \cite{Yukalov_28}. The random 
variables are modeled by a Gaussian white noise \cite{Kampen_29,Kubo_30,Mikhailov_31} 
defined by the stochastic averages
$$
\lgl\lgl \xi_S(t) \rgl\rgl = \lgl\lgl \vp_S(t) \rgl\rgl = 0 \; , \qquad
\lgl\lgl \xi_S(t) \xi_S(t') \rgl\rgl = 2 \gm_3 \dlt( t - t') \; ,
$$
\be
\label{45}
 \lgl\lgl \vp^*_S(t) \vp_S(t') \rgl\rgl = 2 \gm_3 \dlt( t - t') \; , \qquad
\lgl\lgl \xi_S(t) \vp_S(t') \rgl\rgl = \lgl\lgl \vp_S(t) \vp_S(t') \rgl\rgl = 0 \; ,
\ee
where $\gamma_3$ is the attenuation caused by dipolar spin fluctuations. 

Averaging Eqs. (\ref{37}) and (\ref{38}) over spin variables, we take into account 
the existence of longitudinal, $\gamma_1$, and transverse, $\gamma_2$, attenuations 
\cite{Haar_32,Abragam_33}. We define the frequency related to the quadratic Zeeman
effect,
\be
\label{46}  
 \om_Q \equiv (2S - 1) \; \frac{Q}{\hbar} 
\ee
and the effective frequency of spin rotation
\be
\label{47}
\om_s \equiv \om_0 + \om_Q s = \om_0 ( 1 + A s ) \;  ,
\ee
with the dimensionless quadratic Zeeman effect parameter
\be
\label{48}
 A \equiv \frac{\om_Q}{\om_0} = ( 2S - 1 ) \; \frac{Q}{\hbar\om_0} \;  .
\ee
This shows that quadratic Zeeman effect induces an effective anisotropy in the system 
of spinor atoms.

Then we find the equations for the transverse component (\ref{39}),
$$
\frac{du}{dt} = - i (\om_s + \xi_S - i\gm_2 ) u + fs \; + 
$$
\be
\label{49}
 + \; \frac{i}{2}\; (2S -1 )  Q_Z H^2 ( u^* + u) s - \;
\frac{i}{2}\; (2S -1 )  Q_Z B_0 H \left ( w - 2s^2 + u^2 \right ) \; ,
\ee
coherence intensity (\ref{40}),
$$
\frac{dw}{dt} = - 2\gm_2 w + ( u^* f + f^* u ) s \; +
$$
\be
\label{50}
  + \; \frac{i}{2}\; (2S -1 )  Q_Z H^2 \left [ ( u^*)^2 - ( u)^2 \right ] s +
i (2S -1 )  Q_Z B_0 H ( u^* - u ) s^2 \;  ,
\ee
and for the longitudinal spin polarization (\ref{41}), 
$$
\frac{ds}{dt} = - \; \frac{1}{2}\; ( u^* f + f^* u ) \;  -
$$
\be
\label{51}
  - \; \frac{i}{4}\; (2S -1 )  Q_Z H^2 \left [ ( u^*)^2 - ( u)^2 \right ] - \;
\frac{i}{2}\; (2S -1 )  Q_Z B_0 H ( u^* - u ) s -\gm_1 ( s - s_\infty) \; .
\ee
Here $s_\infty$ is a stationary spin polarization and we use the notation
\be
\label{52}
f \equiv - i \left ( \frac{\mu_S H}{\hbar} + \vp_S \right ) \; .
\ee
The transverse attenuation is 
\be
\label{53}
  \gm_2 = \frac{1}{\hbar} \; \rho\mu_F^2 S \; .  
\ee
And the attenuation due to dipolar spin fluctuations can be estimated noticing that 
from equations (\ref{45}) we have
\be
\label{54}
 \gm_3 = \left | \int_0^\infty \lgl\lgl \xi_S(t)\xi_S(0) \rgl\rgl \; dt \right | \;  .
\ee
The equation of motion for $\xi_S(t)$ shows that its time dependence is close to
$$
\xi_S(t) \cong \gm_2 \exp\{ - i(\om_S - i\gm_2)t \} \; .
$$
Therefore, Eq. (\ref{54}) gives the dipolar fluctuation attenuation
\be
\label{55}
 \gm_3 \cong \frac{\gm_2^2}{\sqrt{\gm_2^2 + \om_s^2}} \;  .
\ee

\section{Resonator feedback field}

As is stated in Sec. 3, the external magnetic field (\ref{30}) consists of two terms, 
a constant magnetic field $B_0$ and a magnetic field $H$ created by a magnetic coil 
of an electric circuit. The considered sample of volume $V$ is inserted into the coil 
of volume $V_c$. The electric circuit is characterized by a natural frequency $\omega$ 
and ringing attenuation $\gamma$. The natural frequency $\omega$ is tuned close to the 
Zeeman frequency $\omega_0$, because of which the circuit is called resonant. The moving 
spins of the sample induce in the coil electric current described by the Kirhhoff 
equation. In turn, this current creates a feedback field acting on the spins of the 
sample. The equation for the feedback field follows from the Kirhhoff equation 
\cite{Yukalov_23,Yukalov_24,Yukalov_25,Yukalov_26,Yukalov_27} yielding 
\be
\label{56}
 \frac{dH}{dt} + 2\gm H + \om^2 \int_0^t H(t')\; dt' = 
- 4\pi\eta_c \; \frac{dm_x}{dt} \;  ,
\ee
where $\eta_c \equiv V/V_c$ is the coil filling factor and 
\be
\label{57}
 m_x = \frac{\mu_F}{V} \sum_j \; \lgl S_j^x \rgl  
\ee
is the transverse magnetization density along the coil axis $x$. 

The feedback-field Eq. (\ref{56}) can be rewritten in the integral form    
\be
\label{58}
 H = -4\pi \int_0^t G( t - t') \dot{m}_x(t')\; dt'  ,
\ee
in which the electromotive force is due to the moving magnetization
\be
\label{59}   
\dot{m}_x = \frac{1}{2} \; \eta_c \rho \mu_F S \; \frac{d}{dt}\; (u^* + u)
\ee
and the transfer function is
$$
 G(t) = 
\left [ \cos(\om't) -\; \frac{\gm}{\om'}\; \sin(\om't) \right ] e^{-\gm t} \; ,
\qquad 
\om' \equiv \; \sqrt{\om^2 - \gm^2} \;  .
$$

As usual, we consider the situation when all attenuations are small, such that
\be
\label{60}
 \frac{\gm}{\om} \ll 1 \; , \qquad  \frac{\gm_1}{\om_0} \ll 1 \; , \qquad 
\frac{\gm_2}{\om_0} \ll 1 \; , \qquad  \frac{\gm_3}{\om_0} \ll 1 \; .
\ee
Then the coupling rate, induced by the coupling between the sample and the coil,
\be
\label{61}
 \gm_0 \equiv \frac{\pi}{\hbar}\; \eta_c \rho \mu_F^2 S = \pi \eta_c \gm_2 \; ,
\ee
is also small, as compared to $\omega_0$. 

The integral equation (\ref{58}) can be solved by an iteration procedure, which, to 
first order with respect to the coupling rate (\ref{61}), gives  
\be
\label{62}
  \mu_F H = i\hbar ( u X - X^* u^* ) \; ,
\ee
with the coupling function
\be
\label{63}
 X = \gm_0 \om_s \left [ \frac{1-\exp\{-i(\om-\om_s)t-\gm t\}}{\gm+i(\om-\om_s)} +
\frac{1-\exp\{-i(\om+\om_s)t-\gm t\}}{\gm-i(\om+\om_s)} \right ] \; .
\ee
Here the first term is resonant and prevails over the second, if $\omega_s$ is positive,
while the second term becomes resonant, prevailing over the first, when $\omega_s$ is
negative. Both these cases can be taken into account by the simplified expression
\be
\label{64}
 X \cong \left ( \frac{\gm_0\om_s}{\gm}\right ) \; 
\frac{1-\exp(-i\Dlt_s t-\gm t)}{1 +i\dlt_s} \; ,
\ee
in which
\be
\label{65}
 \Dlt_s \equiv \om - |\om_s| = \om - \om_0 | 1 + As |\; \qquad
\dlt_s \equiv \frac{\Dlt_s}{\gm} \; {\rm sgm}\; \om_s \;  .
\ee
Separating the coupling function into the real and imaginary parts, we define the 
dimensionless coupling functions
\be
\label{66}
 \al \equiv \frac{{\rm Re}X}{\gm_2} \; , \qquad 
\bt \equiv \frac{{\rm Im}X}{\gm_2} \;  .
\ee
Thus we obtain for the real part
\be
\label{67}
\al = \frac{g\gm^2}{\gm^2+\Dlt_s^2}\; ( 1 + As) \left \{
1 - [ \cos(\Dlt_s t) -\dlt_s\sin(\Dlt_s t) ] e^{-\gm t} \right \}
\ee
and for the imaginary part
\be
\label{68}
 \bt = -\; \frac{g\gm^2}{\gm^2+\Dlt_s^2}\; ( 1 + As) \left \{
\dlt_s - [ \sin(\Dlt_s t) + \dlt_s\cos(\Dlt_s t)  ] e^{-\gm t} \right \}\; .
\ee
Here the quantity
\be
\label{69}
g \equiv \frac{\gm_0\om_0}{\gm\gm_2}
\ee
is the dimensionless coupling parameter characterizing the strength of the coupling 
between the sample and the resonant electric circuit.

\section{Averaging of stochastic equations}

Equations (\ref{49}) to (\ref{51}) are stochastic differential equations. Because 
of the existence of small parameters (\ref{60}), it is possible to classify the 
sought quantities into fast and slow functional variables and to treat the equations 
employing averaging techniques \cite{Bogolubov_34}, as applied to stochastic 
differential equations \cite{Freidlin_35}. By the structure of these equations, 
the variable $u$ has to be classified as fast, while $w$ and $s$, as slow. This 
allows us to solve the equation for the fast variable keeping there the slow variables 
as quasi-integrals of motion. 

Substituting the feedback field into Eq. (\ref{49}) yields the equation
\be
\label{70} 
 \frac{du}{dt} = - i\Om u - i\xi_S u - i\vp_S s - X^* u^* s \;  ,
\ee
where
$$
\Om = \om_s - i(\gm_2 - Xs) \; -
$$
\be
\label{71}
  - \;
(2S - 1) \; \frac{Q_Z}{2\mu_F^2} \; \left ( 2 |X|^2 - X^2 \right ) ws - 
i(2S - 1) \; \frac{Q_Z}{2\mu_F} \; B_0 \left ( X w - X^* w - 2X s^2 \right ) \; .
\ee
This equation can be solved keeping the slow variables fixed and treating the 
random variables as functions of time, as is accepted for stochastic equations 
\cite{Morrison_36}. The result is the solution
$$
u = u_0 \exp \left\{ -i \Om t - i \int_0^t \xi_S(t')\; dt' \right\} \; -
$$
\be
\label{72}
 -\;  i s 
\int_0^t \vp_S(t')\exp\left\{ -i\Om(t-t')-i\int_{t'}^t \xi_S(t'')\; dt''\right\}\; dt'\; .
\ee
Then solution (\ref{72}) and the feedback field (\ref{62}) are substituted into 
the equations for the slow variables, which are averaged over time and over the 
stochastic variables. Thus we get the equations for the coherence intensity
$$
\frac{dw}{dt} = - 2\gm_2 w + 2\gm_2 \al ws + 2\gm_3 s^2 \; + 
$$
\be
\label{73}
+ \; 2 (2S-1) Q_Z \; \frac{\hbar\gm_2^2}{\mu_F^2}\; \al\bt w^2 s -
2 (2S-1) Q_Z \; \frac{B_0\gm_2}{\mu_F}\; \al w s^2
\ee
and for the longitudinal spin polarization
$$
\frac{ds}{dt} = - \gm_2 \al w - \gm_3 s \; -
$$
\be
\label{74}
 - \;
(2S-1) \; Q_Z \; \frac{\hbar\gm_2^2}{\mu_F^2}\; \al\bt w^2 +
 (2S-1) Q_Z \; \frac{B_0\gm_2}{\mu_F}\; \al w s - \gm_1(s - s_\infty) \;  .
\ee

Let us introduce the dimensionless quadratic Zeeman-effect parameter
\be
\label{75}
 q \equiv (2S-1) \; \frac{\hbar\gm_2}{\mu_F^2} \; Q_z = S(2S-1)\rho Q_Z \; .
\ee
With the help of this parameter, we can write
$$
(2S-1)\; \frac{B_0}{\mu_F}\; Q_Z = - q \; \frac{\om_0}{\gm_2} \; , \qquad
(2S-1)\; \frac{Q_ZB_0^2}{\hbar\om_0} =  q \; \frac{\om_0}{\gm_2} \; .
$$
And parameter (\ref{48}) becomes
\be
\label{76}
 A = q\; \frac{\om_0}{\gm_2} + \frac{q_Z}{\hbar\om_0} \; .
\ee

Finally, we obtain the equations for the coherence intensity
\be
\label{77}
\frac{dw}{dt} = -2\gm_2 w + 2\gm_2\al w s + 2\gm_3 s^2 + 
2\gm_2 q \al\bt w^2 s + 2\om_0 q \al w s^2
\ee
and for the spin polarization
\be
\label{78}
\frac{ds}{dt} = -\gm_2 \al w - \gm_3 s - 
\gm_2 q \al\bt w^2  - \om_0 q \al w s - \gm_1 ( s - s_\infty ) \;   .
\ee

\section{Analysis of spin dynamics}

Let us first consider the very beginning of the process close to $t = 0$. At very 
short time $t \ra 0$, the coupling functions (\ref{67}) and (\ref{68}) are close to 
zero. The effective spin rotation frequency (\ref{47}) is
$$
\om_s \simeq \om_s(0)  = \om_0 ( 1 + As_0 ) \qquad ( t \ra 0 )
$$
and attenuation (\ref{55}) is
$$
\gm_3 \simeq \frac{\gm_2^2}{\sqrt{\gm_2^2+\om_s^2(0) } } \;  .
$$
When $|A s_0|\ll 1$, then $\om_s\simeq\om_0$ and $\gm_3\simeq\gm_2^2/\om_0$. But if 
$|A s_0|\gg 1$, then $\om_s\simeq\om_0 A s_0$ and $\gm_3\simeq\gm_2^2/\om_0|A s_0|$. 

At the beginning of the process, Eqs. (\ref{77}) and ({78}) simplify to 
\be
\label{79}
 \frac{dw}{dt} = - 2\gm_2 w + 2\gm_3 s^2 \; , \qquad
\frac{ds}{dt} = - \gm_3 s \;  ,
\ee
where we take into account that $\gm_1\ll\gm_3\ll\gm_2$. These equations give
\be
\label{80}
 w \simeq w_0 e^{-2\gm_2 t} + 
\frac{\gm_3s_0^2}{\gm_2-\gm_3} \left ( e^{-2\gm_3 t} - e^{-2\gm_2 t} \right ) \; , 
\qquad s \simeq s_0 e^{-\gm_3 t} \;  ,
\ee
with the initial conditions
$$
 w_0 \equiv w(0) \; , \qquad s_0 \equiv s(0) \;  .
$$

At short time, such that $\gamma_2 t \ll 1$, we get
\be
\label{81}
 w \simeq w_0(1-2\gm_2t) + 2s_0^2\gm_3 t \; , \qquad
s \simeq s_0(1-\gm_3 t) \qquad ( \gm_2 t \ll 1 ) \;  .
\ee
The form of these solutions shows the importance of spin fluctuations responsible 
for the appearance of the attenuation $\gamma_3$. Such spin fluctuations play the 
role of a trigger starting spin motion. The standard semiclassical approximation, 
where spin fluctuations are not taken into account, hence $\gamma_3$ is set to zero, 
would not lead to noticeable spin relaxation, if no initial coherence is imposed on 
the sample, hence if $w_0=0$, but could only exhibit a very slow relaxation of the 
spin polarization to $s_\infty$ during rather long time $T_1\equiv 1/\gm_1$. 

At arbitrary time, we need to solve Eqs. (\ref{77}) and ({78}) numerically. For 
this purpose, it is convenient to measure time in units of $1/\gamma_2$. Also, we 
consider the case of resonance $\om_0=\om$, when the Zeeman frequency coincides with 
the circuit natural frequency. Then the equations to be solved acquire the form
\be
\label{82}
 \frac{dw}{dt} = - 2w + 2\al ws + 2 \; \frac{\gm_3}{\gm_2}\; s^2 + 
2q\al\bt w^2 s + 2q \; \frac{\om}{\gm_2}\; \al w s^2 \;  ,
\ee
and
\be
\label{83}
\frac{ds}{dt} = - \al w - \; \frac{\gm_3}{\gm_2}\; s - 
q\al\bt w^2 - q \; \frac{\om}{\gm_2}\; \al w s \;  ,
\ee
in which $\alpha$ and $\beta$ are the coupling functions (\ref{67}) and (\ref{68}), 
while $\gm_3$ is attenuation (\ref{55}).  

In order to estimate the typical parameters of a spinor atomic system, let us take 
the values corresponding to spinor Bose atoms with $F=1$, such as $^7$Li, $^{23}$Na, 
$^{41}$K, and $^{87}$Rb (see \cite{Stamper_7,Yukalov_9,Frapolli_37} and references 
there in). The hyperfine splitting energy \cite{Grigoriev_38} is of the order 
$\Dlt W\sim 10^{10}\hbar/{\rm s}\sim 10^{-17}$ erg. Then $Q_Z\sim 10^{-24}{\rm cm}^3$.
For the atomic density $\rho \sim 10^{15}{\rm cm}^{-3}$, we have 
$q\sim\rho Q_Z\sim 10^{-9}$. Taking $\mu_F\sim\mu_B$ gives $\rho\mu_F^2\sim 10^{-25}$ erg. 
Therefore $\gamma_2 \sim 10^2 {\rm s}^{-1}$. The Zeeman frequency 
$\om_0\sim\mu_F B_0/\hbar\sim 10^7 B_0/$ Gs depends on the external magnetic field. 
For $B_0 \sim (1 - 10^4)$ G, we get $\omega_0 \sim (10^7 - 10^{11}) {\rm s}^{-1}$. 
Hence $\om_0/ \gm_2 \sim 10^5 - 10^9$. We take into account that $\gm_0 \sim \gm_2$,
$g \sim \omega_0/\gamma$, and $\alpha \sim \beta \sim g \sim \omega_0/ \gamma$. 
Thus $q \omega_0/\gamma_2 \sim 10^{-11} \omega_0 {\rm s} \sim 10^{-4} - 1$. The 
alternating-current Zeeman effect parameter can reach $q_Z \sim 10^5 \hbar/{\rm s}$, 
from where $q_Z/\hbar\om_0\sim 10^{-6}-10^{-2}$. Therefore 
$A\simeq q\om_0/\gm_2\sim 10^{-4}-1$. When the Zeeman frequency is in resonance 
with the electric circuit natural frequency, $\omega_0 = \omega$, then $|A s|$ can 
be of order of one or smaller and the effective detuning defined in Eq. (\ref{65}) 
is
\be
\label{84}
 \Dlt_s = - \om_0 As = - q \; \frac{\om_0^2}{\gm_2} \; s \;  .
\ee
Generally, the parameter $A$, due to quadratic Zeeman effect, can be either 
positive or negative.    

Solving Eqs. (\ref{82}) and (\ref{83}), we are interested in a self-organized 
process, when the spin motion is not pushed by an externally imposed coherent 
field, but starts from natural spin fluctuations inside the system and develops 
through the nonlinear interaction with the resonator feedback field. This implies 
the initial condition $w_0 = 0$. 

For the longitudinal spin polarization, we take the initial condition $s_0 = 1$. 
This corresponds to a strongly nonequilibrium situation. The case of an equilibrium 
initial condition, when $s_0$ equals minus one, is not interesting, since then the 
system stays in the given state, just slightly oscillating around it and exhibiting 
no nontrivial dynamics.      

From Eqs. (\ref{82}) and (\ref{83}) it is seen that spin dynamics strongly depends 
on the coupling functions (\ref{67}) and (\ref{68}) that are proportional to the 
effective coupling parameter
\be
\label{85}
 g_{eff} \equiv \frac{g\gm^2}{\gm^2 + \Dlt_s^2} \; ( 1 + As_0 ) \; .
\ee
Substituting here the expressions for $g$, $A$, and $\Delta_s$ yields
\be
\label{86}
 g_{eff} \equiv \left ( \frac{\gm\gm_0}{\gm_2^2}\right ) \;
\frac{(1+qs_0\om/\gm_2)\;\om/\gm_2}{(\gm/\gm_2)^2+q^2s_0^2(\om/\gm_2)^4} \; .
\ee

The behavior of the parameter $g_{eff}$ as a function of the frequency $\om$ is 
essentially influenced by the value of the quadratic Zeeman effect parameter $q$.  
Respectively, since the occurrence of noticeable spin motion is governed by the 
coupling functions, the delay time, when such a motion starts is proportional to 
the inverse of $g_{eff}$. The larger $g_{eff}$, the shorter the delay time. In 
the absence of the quadratic Zeeman effect, when $q = 0$, we have
\be
\label{87}
 g_{eff} = \frac{\gm_0\om}{\gm\gm_2} = g \qquad ( q = 0 ) \;  .
\ee
Then $g_{eff}$ monotonically increases with the increase of $\omega$, hence the 
delay time diminishes.

But in the presence of the quadratic Zeeman effect, the behavior of $g_{eff}$ 
is not monotonic. At small $\omega$, we get
\be
\label{88}
 g_{eff} \simeq \frac{\gm_0\om}{\gm\gm_2}  \qquad ( \om \ra 0 ) \;  .
\ee
So that the coupling parameter first increases with $\omega$, and the delay 
time diminishes. But at large $\omega$, the behavior is different:
\be
\label{89}
  g_{eff} \simeq \frac{\gm\gm_0}{qs_0\om}  \qquad ( \om \ra \infty ) \;  .
\ee
Then the magnitude of the coupling parameter diminishes with $\omega$, hence the 
delay time increases. The change of the behavior happens when $\omega$ is close 
to the critical value
\be
\label{90}
 \om_c \sim \sqrt{ \frac{\gm\gm_2}{|qs_0| } } \;  .
\ee
More precisely, the maximum of the coupling parameter (\ref{88}) is given by the 
solution to the equation
$$
 2q^3 \om^5 + 3q^2\om^4 - 2q\gm^2\om - \gm^2 = 0 \;  ,
$$
where we set $s_0 = 1$ and $\omega$ and $\gamma$ are measured in units of $\gm_2$.  
  
The influence of the quadratic Zeeman effect is illustrated by the numerical solution 
of Eqs. (\ref{82}) and (\ref{83}). When the quadratic Zeeman effect is absent, hence 
$q = 0$, the spin polarization $s$ and coherence intensity $w$ are shown in Fig. 1. 
In agreement with the above discussion, the delay time of spin reversal diminishes 
with increasing $\omega$. The spin reversals are accompanied by the pulses of 
coherence intensity. The coherent motion of spins develops due to the action of 
the feedback field.  

At small, but finite, parameters $q$ and for the frequencies smaller than the critical 
frequency, the behavior of solutions is similar to the case of zero $q$. However, 
approaching the critical frequency, the solutions exhibit oscillations after the 
spin reversal, which is caused by the oscillations in the coupling functions. This 
is demonstrated in Fig. 2, where for $\om=10^3$ and $\om=10^4$, there occurs a complete 
spin reversal without oscillations. And for $\om=10^5$, the spin reversal is slightly 
incomplete and there are oscillations after the reversal. There is no noticeable 
difference for positive or negative $q=\pm 10^{-8}$. Here and in what follows, the 
frequencies and attenuations are measured in units of $\gamma_2$. Spin reversal time 
diminishes with increasing $\omega$. 

For $q = 10^8$ and the frequencies larger than $\omega = 10^5$, the spin reversal 
time increases with increasing $\omega$, as is explained above and as is shown in 
Fig. 3.

In the vicinity of the critical frequency, there appears the dependence on the sign 
of $q$, as is illustrated in Fig. 4. The spin reversal time is larger for the negative 
$q$, since, as is seen from Eq. (\ref{85}), a negative $A$ diminishes the effective 
coupling parameter $g_{eff}$, hence increases the spin reversal time. 

As is demonstrated in Fig. 2, the oscillations in the solutions appear at large 
frequencies. To realize spin reversal without oscillations, one can proceed as 
follows. Notice that the oscillations arise due to the oscillating behavior of 
the coupling functions (\ref{67}) and (\ref{68}), which is caused by a nonzero value 
of the effective detuning $\Delta_s$ defined in Eq. (\ref{65}). It is clear that to 
make oscillations smaller, it is necessary to diminish the parameter $A$, given in 
Eq. (\ref{76}). This parameter combines the terms due to both the static-current 
quadratic Zeeman effect and alternating-current quadratic Zeeman effect. Since the 
sign of the alternating-current quadratic Zeeman effect parameter $q_Z$, defined in 
Eq. (\ref{9}), can be easily varied by varying the detuning $\Delta$, it is possible 
to make its sign opposite to that of the static-current quadratic Zeeman effect 
parameter $Q_Z$, given in Eq. (\ref{8}). This reduces the value of the combined 
parameter $A$, as a result reducing oscillations of the coupling functions (\ref{67}) 
and (\ref{68}). This compensation effect is illustrated in Fig. 5 for small $|A|\ll 1$. 
Then spin oscillations are suppressed for all frequencies and spin reversal time 
decreases with increasing $\omega$.

\section{Conclusion}

We have considered a system of spinor atoms loaded into a deep optical lattice, where 
atomic degrees of motion are frozen, while effective spin variables can move, being 
regulated by external magnetic fields. The sample is subject to a static magnetic 
field and a feedback field of a magnetic coil of a resonance electric circuit. The 
existence of the feedback field makes it possible to realize a coherent motion of spins 
leading to a fast spin reversal, when the sample is initially prepared in a strongly 
nonequilibrium state. 

A system of spinor atoms, in addition to the usual linear Zeeman effect, can experience 
two types of quadratic Zeeman effect, the so-called nonresonant static-current quadratic 
Zeeman effect and a quasi-resonant alternating-current quadratic Zeeman effect. The 
influence of both these effects on spin dynamics is studied. Conditions are emphasized, 
when it is possible to realize fast spin reversal from an initially prepared strongly 
nonequilibrium state. Such spin reversals can find wide applications in spintronics and 
information processing.     

For concreteness, we have considered spinor atoms, but, we think, the application 
of the present theory can be much wider. For instance, quantum dots in many aspects 
are similar to atoms, often being even termed {\it artificial atoms} \cite{Birman_39}. 
Quantum dots can also experience quadratic Zeeman effect \cite{Prado_40}. A system of 
quantum dots could be another example of a nontrivial influence of quadratic Zeeman 
effect on spin dynamics.    

\vskip 3mm

{\it Author contribution statement}

\vskip 2mm

All authors contributed equally to the paper.

\newpage

\begin{center}

{\Large{\bf Figure Captions}}

\end{center}

\vskip 1cm
{\bf Figure 1}. Temporal behavior of the longitudinal spin polarization $s$ (a) 
and coherence intensity $w$ (b) for $q = 0$, $\gamma = 10$, and different $\om$ 
measured in units of $\gamma_2$. Time is measured in units of $1/ \gamma_2$. The 
delay time of spin reversals diminishes with increasing $\omega$.  

\vskip 1cm
{\bf Figure 2}. Time dependence of the spin polarization $s$ (a) and coherence 
intensity $w$ (b) for $\gamma = 10$, $q = \pm 10^{-8}$, and different $\omega$ 
in units of $\gamma_2$. Time is measured in units of $1/ \gamma_2$. For the 
frequencies smaller than $\om=10^5$, the time of spin reversal diminishes with 
increasing $\om$.  

\vskip 1cm
{\bf Figure 3}. Spin polarization $s$ (a) and coherence intensity $w$ (b) as 
functions of time, measured in units of $1/ \gm_2$, for $q=10^{-8}$, $\gm=10$ and 
different frequencies larger than or equal to $\om=10^5$, in units of $\gm_2$. 
Spin reversal time increases with increasing $\om$.    

\vskip 1cm
{\bf Figure 4}. Spin polarization $s$ (a) and coherence intensity $w$ (b) 
as functions of time, measured in units of $1/ \gamma_2$, for $\gamma=10$, 
$\om=1.394\cdot 10^5$, in units of $\gm_2$, and for $q=\pm 10^{-8}$.

\vskip 1cm
{\bf Figure 5}. Longitudinal spin polarization as a function of time measured in 
units of $1/\gm_2$ for $|A|\ll 1$, $\gm=10$, $q=\pm 10^{-7}$ and different $\om$ 
measured in units of $\gm_2$. Spin reversal time decreases with increasing $\om$.

\newpage

\begin{figure}[ht]
\centerline{\hbox{
\includegraphics[width=7.4cm]{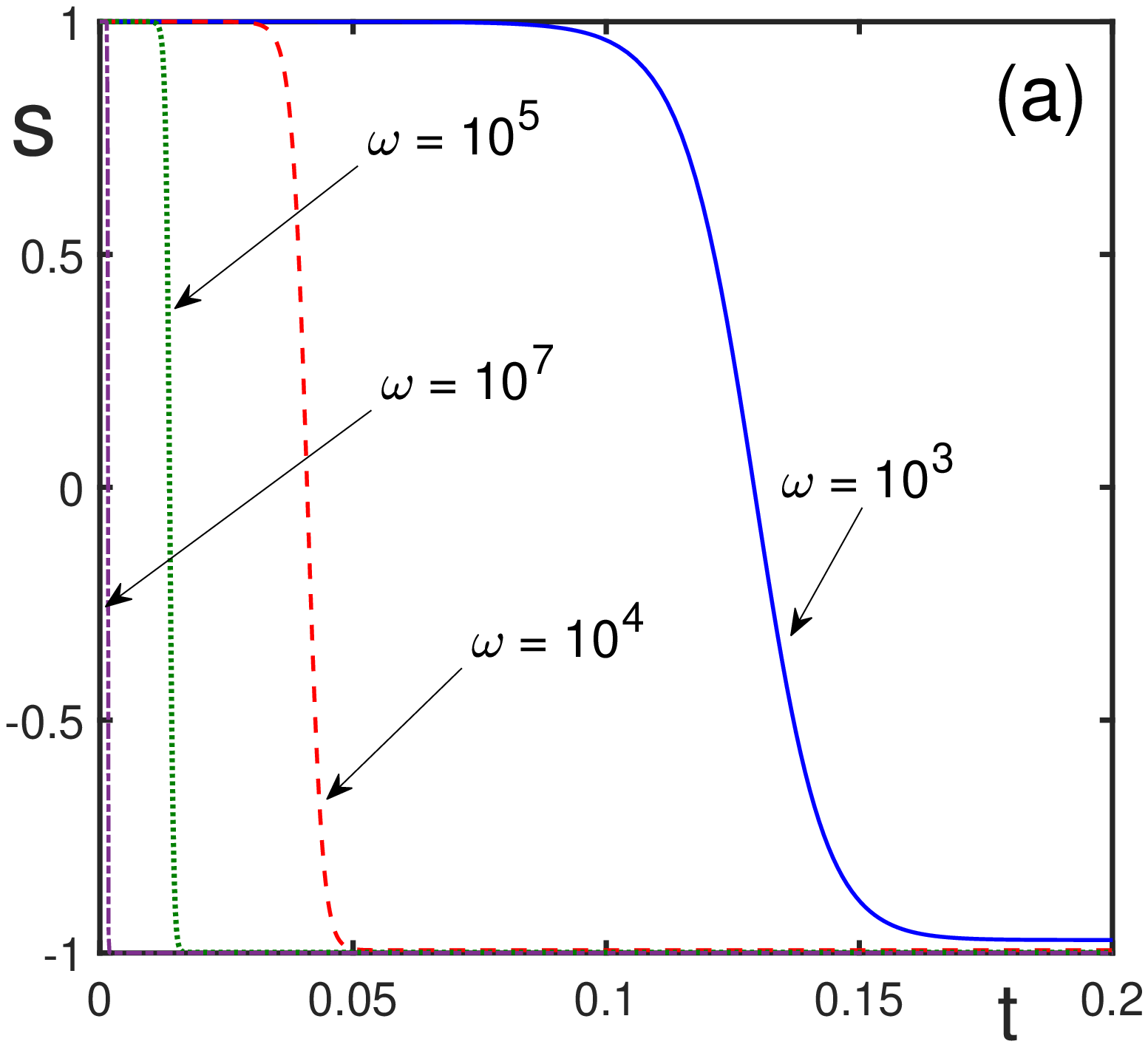} \hspace{2cm}
\includegraphics[width=7.4cm]{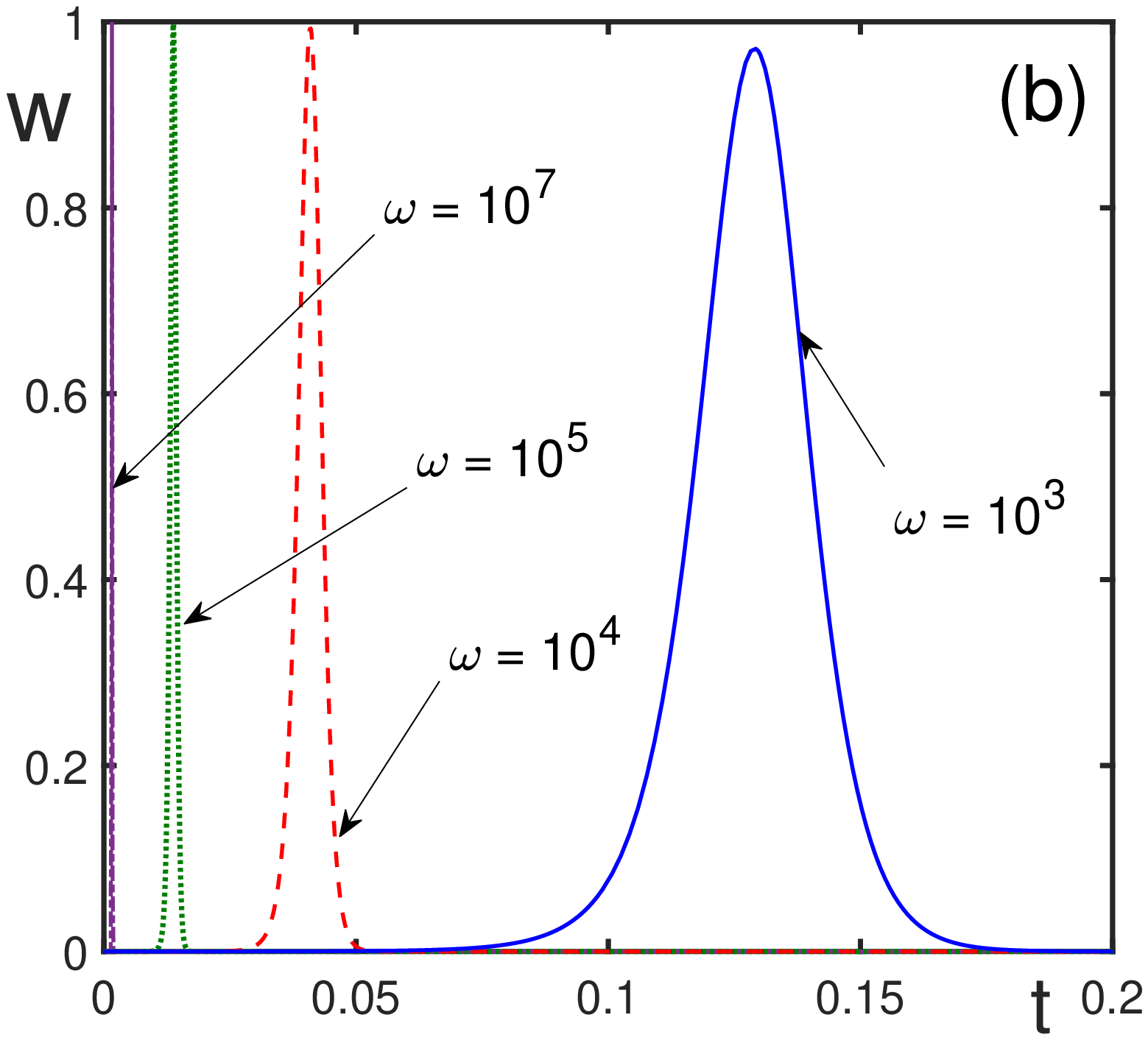} }   }
\caption{Temporal behavior of the longitudinal spin polarization $s$ (a) 
and coherence intensity $w$ (b) for $q = 0$, $\gamma = 10$, and different $\om$ 
measured in units of $\gamma_2$. Time is measured in units of $1/ \gamma_2$. The 
delay time of spin reversals diminishes with increasing $\omega$.
}
\label{fig:Fig.1}
\end{figure}

\vskip 2cm

\begin{figure}[ht]
\centerline{\hbox{
\includegraphics[width=7.4cm]{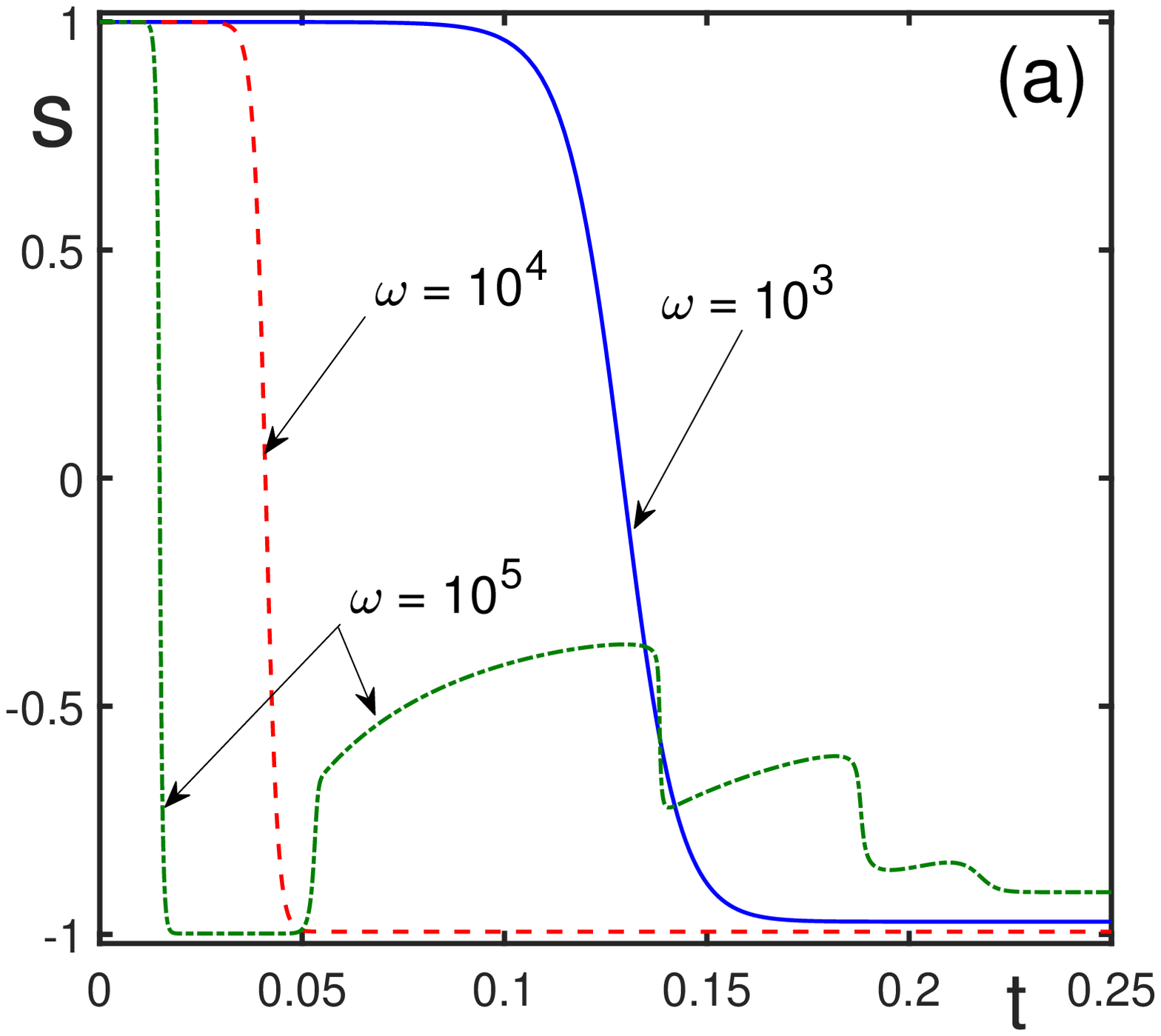} \hspace{2cm}
\includegraphics[width=7.4cm]{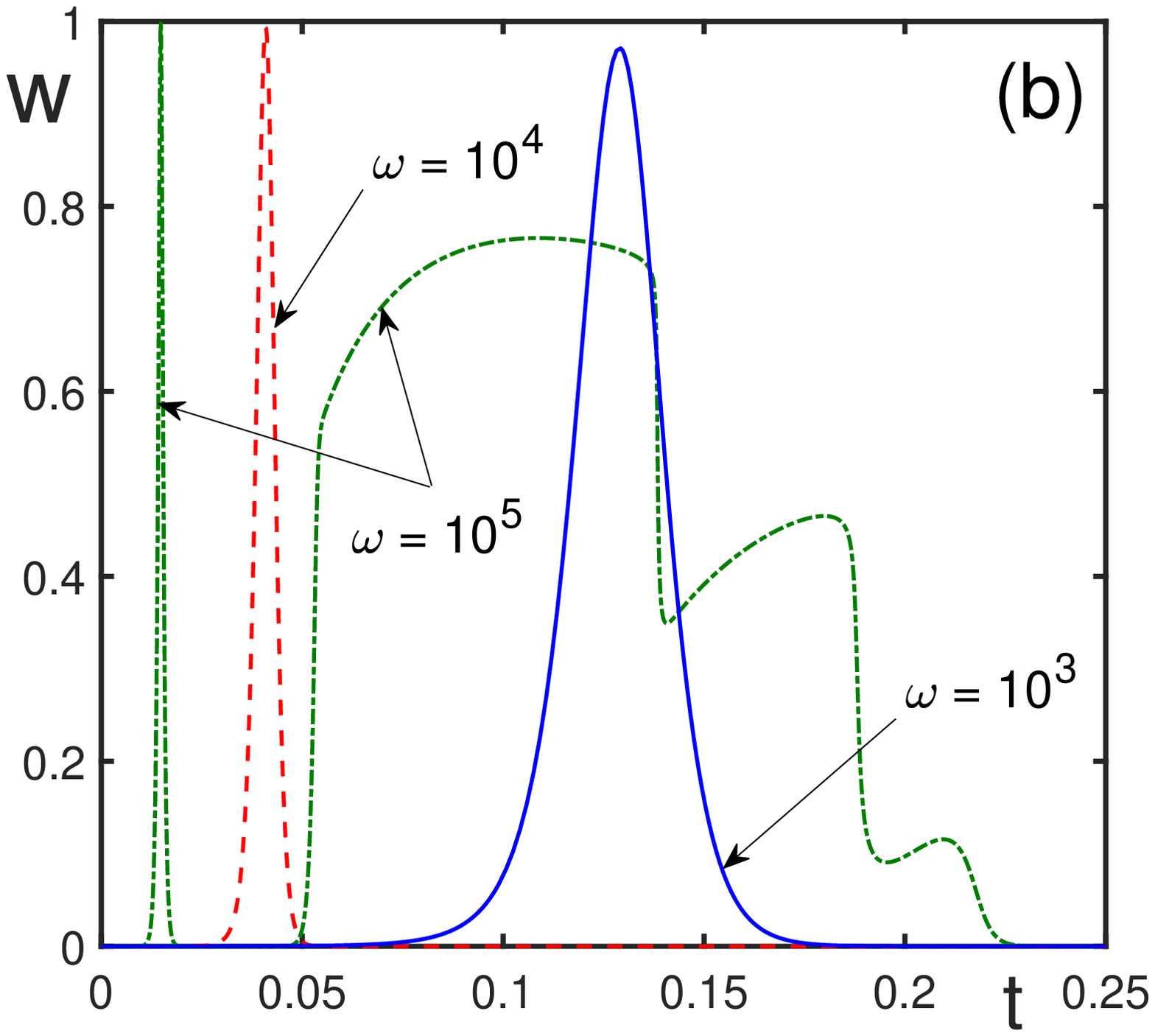} }   }
\caption{Time dependence of the spin polarization $s$ (a) and coherence 
intensity $w$ (b) for $\gamma = 10$, $q = \pm 10^{-8}$, and different $\omega$ 
in units of $\gamma_2$. Time is measured in units of $1/ \gamma_2$. For the 
frequencies smaller than $\om=10^5$, the time of spin reversal diminishes with 
increasing $\om$.
}
\label{fig:Fig.2}
\end{figure}

\begin{figure}[ht]
\centerline{\hbox{
\includegraphics[width=7.4cm]{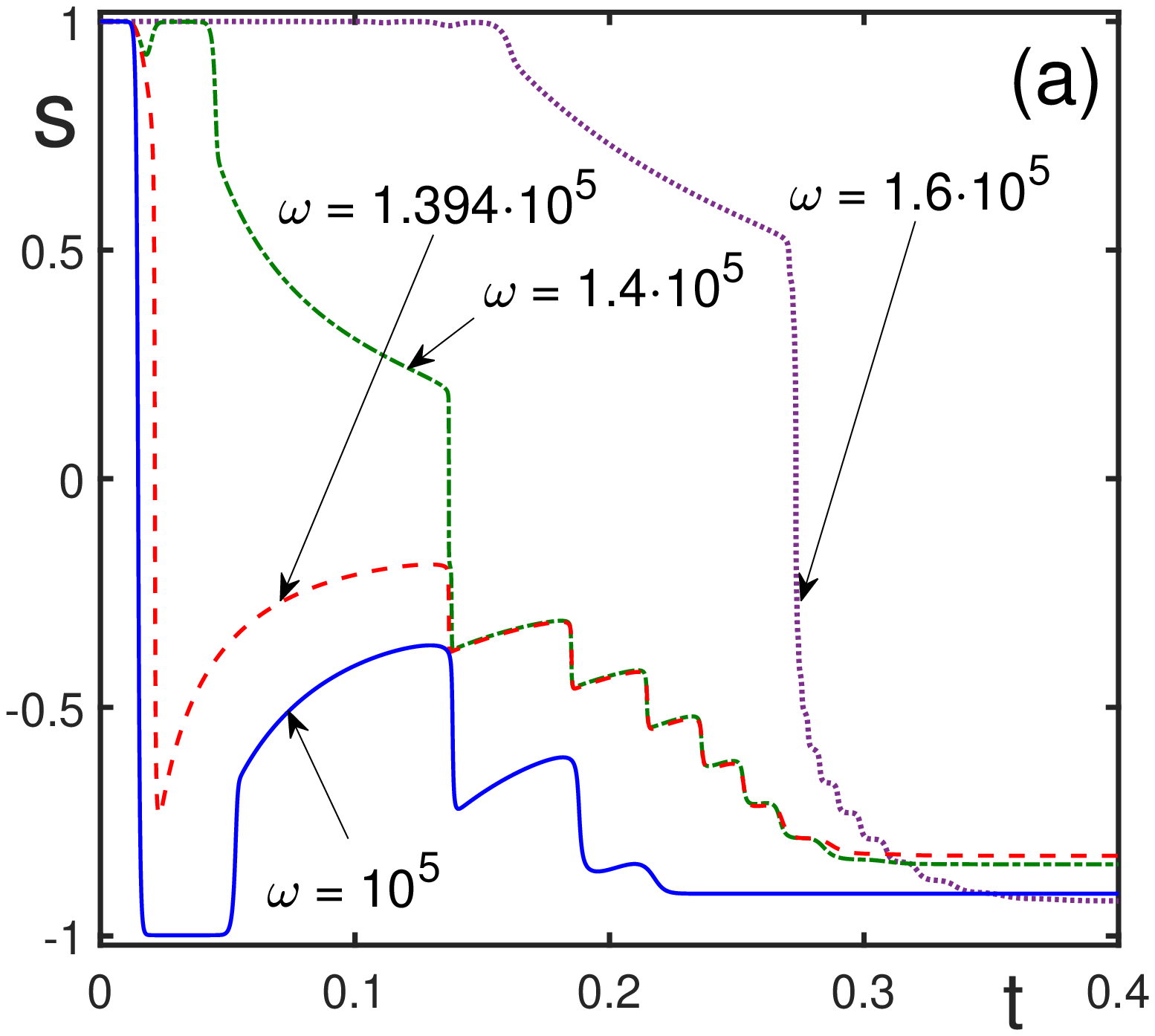} \hspace{2cm}
\includegraphics[width=7.4cm]{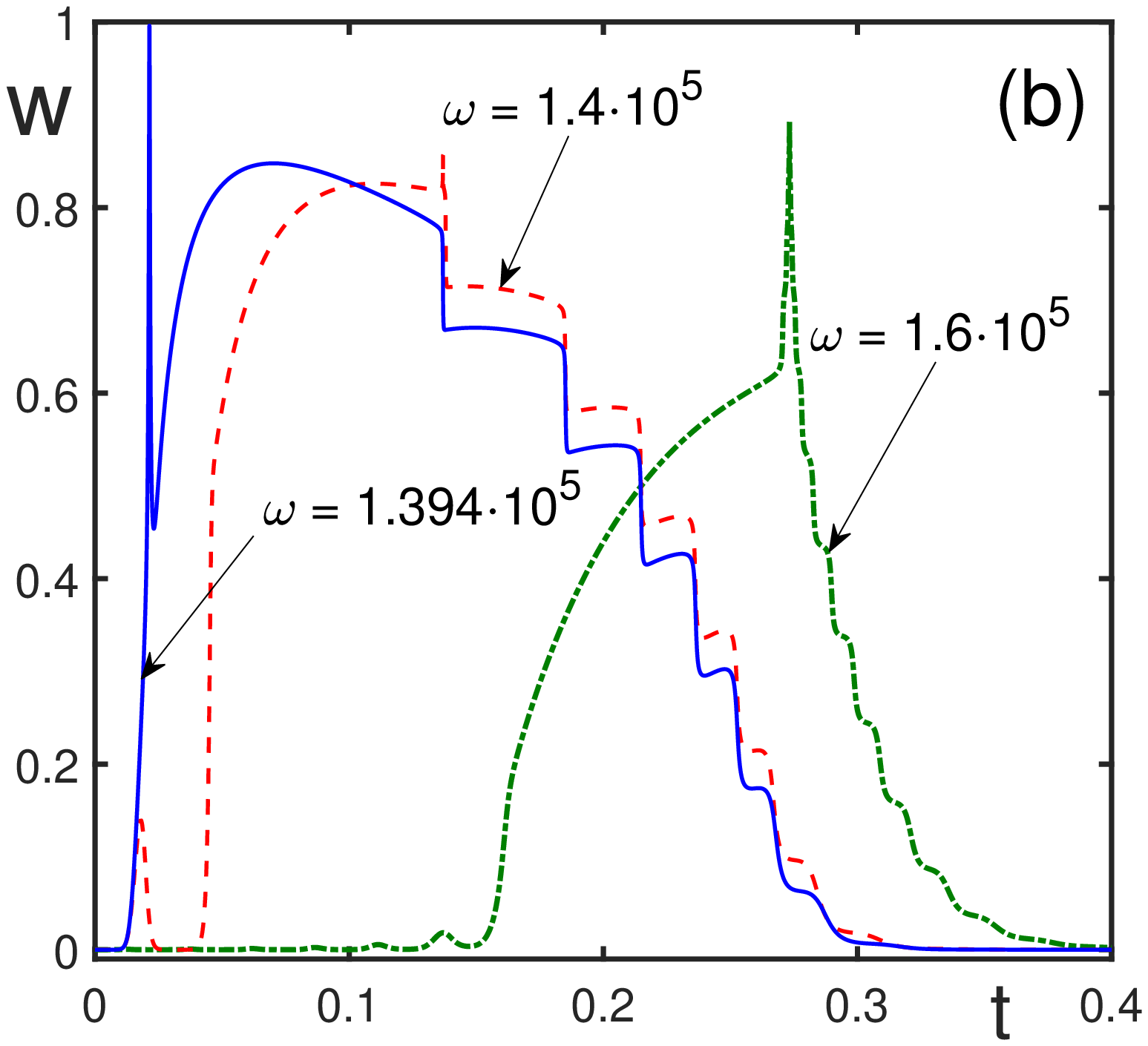} }   }
\caption{Spin polarization $s$ (a) and coherence intensity $w$ (b) as 
functions of time, measured in units of $1/ \gm_2$, for $q=10^{-8}$, $\gm=10$ and 
different frequencies larger than or equal to $\om=10^5$, in units of $\gm_2$. 
Spin reversal time increases with increasing $\om$.
}
\label{fig:Fig.3}
\end{figure}

\vskip 2cm

\begin{figure}[ht]
\centerline{\hbox{
\includegraphics[width=7.4cm]{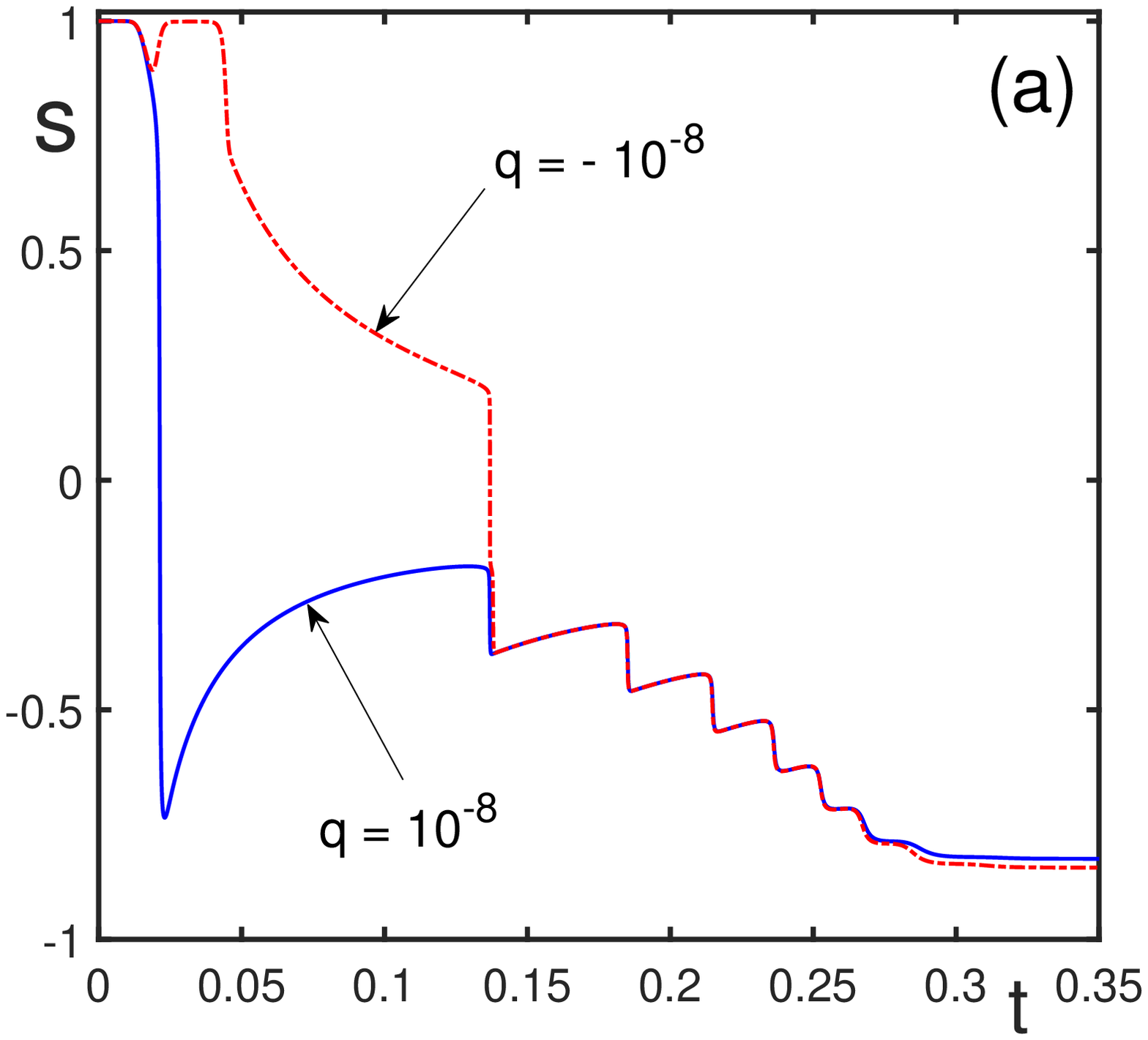} \hspace{2cm}
\includegraphics[width=7.4cm]{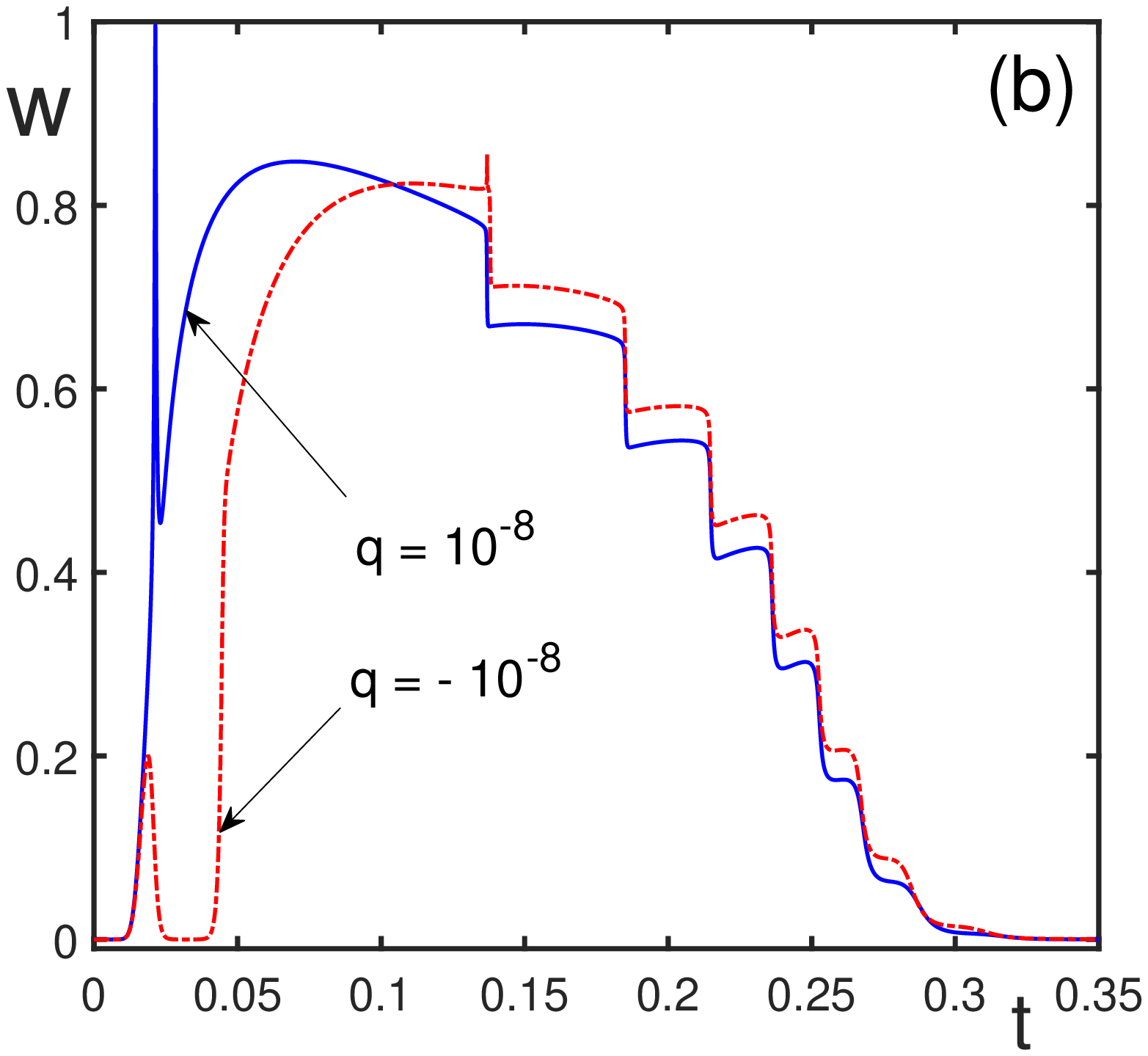} }   }
\caption{Spin polarization $s$ (a) and coherence intensity $w$ (b) 
as functions of time, measured in units of $1/ \gamma_2$, for $\gamma=10$, 
$\om=1.394\cdot 10^5$, in units of $\gm_2$, and for $q=\pm 10^{-8}$. 
}
\label{fig:Fig.4}
\end{figure}

\begin{figure}[ht]
\centerline{
\includegraphics[width=7.5cm]{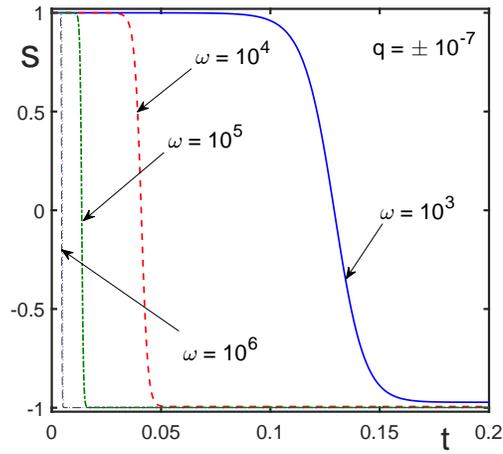} 
}
\caption{Longitudinal spin polarization as a function of time measured in 
units of $1/\gm_2$ for $|A|\ll 1$, $\gm=10$, $q=\pm 10^{-7}$ and different $\om$ 
measured in units of $\gm_2$. Spin reversal time decreases with increasing $\om$.
}
\label{fig:Fig.5}
\end{figure}

\end{document}